\documentclass[prl,showpacs,twocolumn]{revtex4}
\usepackage{amsfonts}
\usepackage{amsmath}
\usepackage{amssymb}
\usepackage{graphicx}

\begin{document}

\title{Photovoltaic Effect of Atomtronics Induced by Artificial Gauge Field}
\author{Wenxi Lai$^{1,2}$,Yuquan Ma$^{1}$, and W. M. Liu$^{2}$}\email{wmliu@iphy.ac.cn}

\affiliation{1 Beijing Information Science and Technology University, Beijing 100192, China}
\affiliation{2 Institute of Physics, Chinese Academy of Sciences, Beijing 100190, China}

\begin{abstract}
We investigate photovoltaic effect of atomtronics induced by artificial gauge field in four optical potentials. Effective magnetic flux gives rise to polarization of atom occupation probability which creates current of atomtronics. The relation between atomic current and magnetic flux behaves like the current-phase property in Josephson junction. The photovoltaic cell is well defined by the atomic opened system which have effective voltage and two different poles that correspond to two internal states of atomtronics. The atom flow is controllable by changing the direction of incident light and other system parameters. Detection of the atomic current intensity is available through light emission optical spectrum in experiments.
\end{abstract}

\maketitle

Photovoltaic transistors of electrons are widely studied in semiconductors~\cite{Ganichev2001,Oka,Ganichev2010,Delerue,Semonin,Bonaccorso,Chan,Taguchi}. To enhance the efficiency of such transistors, there is a tendency that electronic components are manufactured smaller and smaller in size. When the size of these devices becomes very tiny, electrons would hard to be detectable and their heat effect becomes serious. In contrast, atoms are very active in light and have much large volume comparing with electrons, the former appear more advantages in designing future devices~\cite{Chin}. Atomic transistors are widely studied recently by creating non-equilibrium states of atomic gas~\cite{Daley,Fuechsle,Tettamanzi,Xie}. However, at least two basic conditions should be necessary for a photovoltaic atomic transistor. One is coupling between momentum of atoms and applied light, the other is an opened system of atoms.

Momentum can transmit from electromagnetic field to atoms through spin-orbit coupling~\cite{Seaman,Ramanathan,Lee,Eckel,Zhang,CYLai,Li,Haug}. The effect of spin-orbit coupling in atomic system has been developed recently using artificial gauge field and interesting phenomenons are exploited such as effective magnetic flux in synthetic dimensions of atoms~\cite{Livi,Kolkowitz,Wall}, quantum spin Hall effect~\cite{Kennedy,Aidelsburger,Stuhl}, chiral conductors ~\cite{Kollath,Mancini,An}, superradiance induced particle flow~\cite{Zheng}, Kondo effect~\cite{Jiang,Bauer}. In the above phenomenons, dynamics of atoms play like charged particles in electromagnetic fields~\cite{Juzeliunas2010,Gerbier,Dalibard,Goldman2014,Zhang2012,Su2016,Liu2017}. The earlier works reveal that atoms perform differently in special motion depending on their internal states in artificial gauge fields.

In this Letter, we combine the electrodynamic feature of atoms in artificial gauge field and local optical potentials to propose a model of atomic photovoltaic cell. Similar to electrons in semiconductor quantum dots, cold atoms can be trapped in a particular optical wells~\cite{Recati,Knap}. Quite recently, the opened system of cold atoms has been realized in experiments based on the optical wells~\cite{Caliga1,Caliga2}, where nonequilibrium atom flow has been observed under different chemical potentials. In order to create net current we need at least two atomic quantum dots to grantee that the phase of atom wave function coherently amplified when the atom moves from one dot to the other dot under optical driving. Gradient of the phase (scalar) through displacement represents the artificial gauge field. Therefore, coherence of atom-light interaction is the key mechanism of the artificial gauge field in cold atoms.

\begin{figure}
\includegraphics[width=8cm]{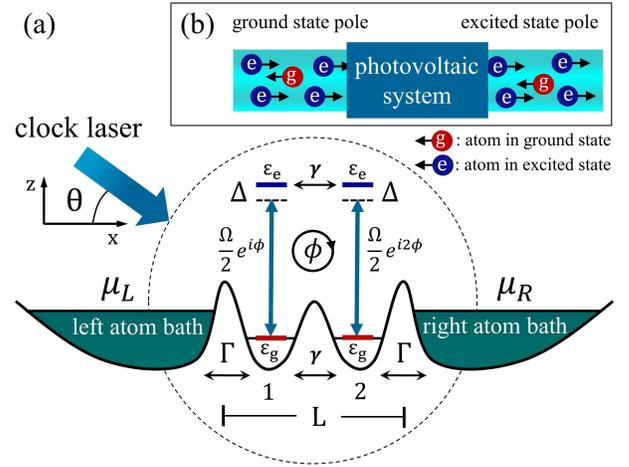}\
\caption{(Color on line) (a) Atom baths are located on the two sides of the atomic quantum dots with the same chemical potentials $\mu$. $x$ axes is parallel to the array of four optical wells. A clock laser is acting the two dots with angle $\theta$ from the $x$ axes. In this case, the clock laser gives momentum $\Delta k=\frac{2\pi cos\theta}{\lambda_{C}}$ to each atom. Due to the spin orbit coupling effect, a net artificial magnetic flux $\phi=\Delta k L$ occurs in area rounded by a closed trajectory in the synthetic two dimension. (b) A schematic picture of the photovoltaic cell in figure (a) is shown here. It has two `electrode` related to the ground state and excited state atoms. Each arrow shows direction of atom motion.}\label{sys}
\end{figure}

The basic setup is shown in Fig.~\ref{sys}. In the four optical wells, two of them are the atomic quantum dots with narrow potentials and play the role of photovoltaic system. The other two work as conductors that connect to the photovoltaic system on the two sides, respectively. We use Fermion atoms and each dot is assumed to allow only one atom occupation due to Pauli's exclusion principle. The rest two wells localized on left and right side are very wide to form source and drain with large number of atoms. Atoms can tunnel through the barrier between neighboring optical wells. A synthetic dimension is formed due to optical transition between internal atom states and atom tunneling between two neighboring quantum dots. We consider rare earth atom $^{173}Yb$ which have a ground state $g=$ $^{1}S_{0}$ and a matastable state $e=$ $^{3}P_{0}$, supporting optical clock transition with a coherent life-time of $20$ s. A clock laser at wave length $\lambda_{C}=578$ nm is acting on the atomic quantum dots. The angle between incident clock laser and the line of optical well array is $\theta$.

Coherent process is happened in the two quantum dots system and phase change of optical transition in the left dot is assumed to be $\phi$, relatively, in the right dot it is $2 \phi$ ~\cite{Livi}. The phase difference $\phi=2\pi L \cos\theta/\lambda_{C}$ is comes from momentum change $\Delta k=2\pi cos\theta/\lambda_{C}$ of an atom during clock field transition and the distance of the atom tunneling $L$. $L$ is distance between the two atomic quantum dots. The net phase $\phi$ represent artificial magnetic field flux that induced by the coherent clock transition of atoms. Direction $\theta$ of the clock laser can be tuned to change the effective magnetic flux and allow one to control the atomtronic transistor.

Hamiltonian of the photovoltiac transistor can be written in three parts as $H = H_{S}+H_{B}+H_{I}$. The first term is Hamiltonian of the two atomic quantum dots driven by the clock laser with frequency $\omega_{c}$,
\begin{eqnarray}
H_{S} &=& \sum_{j=1,2;s=g,e}\varepsilon_{s} a_{j,s}^{\dag}a_{j,s}+\hbar\gamma (\sum_{s=g,e}a_{1,s}^{\dag}a_{2,s}+h.c.) \notag \\ &&+\frac{\hbar\Omega}{2}\sum_{j=1,2}(e^{i\omega_{c} t} e^{i j \phi} a_{j,g}^{\dag}a_{j,e}+h.c.),
\end{eqnarray}
where, $a_{j,s}$ ($a_{j,s}^{\dag}$) is annihilation (creation) operator of atoms in the $j$th quantum dot. The atomtronics are characterized by two internal states with ground level $\varepsilon_{g}$ and excited level $\varepsilon_{e}$. The ground state energy levels of both the two quantum dots are assumed to be zero. Atom tunneling between the two dots occurs with the transition rate $\gamma$. $\hbar$ is the Planck constant. A clock light is driving single atoms with Rabi frequency $\Omega$. Relative phase change along with the atom transition is described by the term $j \phi$ at position $j$. The frequency of optical field is very fast comparing with the Rabi frequency, therefore we take the rotating wave approximation for the atom-field interactions. The atomic leads are described by the Hamiltonian of free atomic gas
\begin{eqnarray}
H_{B} &=&\sum_{\alpha=L,R;s=g,e;k} \varepsilon_{k} b_{\alpha,s,k}^{\dag}b_{\alpha,s,k} .
\end{eqnarray}
Here, $b_{\alpha,s,k}$ ($b_{\alpha,s,k}^{\dag}$) is annihilation (creation) operator of the atoms with energy $\varepsilon_{k}$ in the atom leads $\alpha=L,R$. The atomic bath can be treated as cold atom clouds.

The couplings between the system and the atom gases are written as
\begin{eqnarray}
H_{I} &=& \hbar g \sum_{s=g,e;k} ( b_{L,k,s}^{\dag}a_{1,s}+ b_{R,k,s}^{\dag}a_{2,s}+h.c.).
\end{eqnarray}
Atoms are in or out from the atom bath with tunneling amplitude $g$.

Using the Markovian approximation to the coupling between system and atom conductors, we obtain the following master equation for the atom-light opened system~\cite{Wenxi,Walls,Scully},

\begin{eqnarray}
\frac{\partial}{\partial t}\rho &=&-\frac{i}{\hbar} [H^{eff}_{S},\rho]+\mathcal{L}_{L}\rho+\mathcal{L}_{R}\rho.
\label{equation}\end{eqnarray}
The first term on the right side of Eq.\eqref{equation} represents evolution of the double-dot system with the effective Hamiltonian $H^{eff}_{S}=\sum_{j=1,2}\Delta a_{j,e}^{\dag}a_{j,e}+\frac{\hbar\Omega}{2}\sum_{j=1,2}(e^{i j\phi} a_{j,g}^{\dag}a_{j,e}+h.c.)+\hbar\gamma \sum_{s=g,e}(a_{1,s}^{\dag}a_{2,s}+h.c.)$, where $\Delta=\varepsilon_{e}-\varepsilon_{g}-\hbar \omega_{c}$ describes the detuning between the clock field and atoms. The rest terms of Eq.\eqref{equation} can be written as $\mathcal{L}_{\alpha}\rho=\frac{\Gamma}{2}\sum_{s=g,e}[f_{\alpha}(\varepsilon_{s})(2a_{j,s}^{\dag}\rho a_{j,s}-\{a_{j,s}a_{j,s}^{\dag},\rho\})+(1-f_{\alpha}(\varepsilon_{s}))(2a_{j,s}\rho a_{j,s}^{\dag}-\{a_{j,s}^{\dag}a_{j,s},\rho\})]$ with the anti-commutation relation $\{O,\rho\}$ for any operator $O$. $j=1$ ($j=2$) when $\alpha=L$ ($\alpha=R$). The Liouville super-operators $\mathcal{L}_{L}$ and $\mathcal{L}_{R}$ acting on the density matrix $\rho$ coupling between the double-quantum-dot and two atomic baths with coupling strength $\Gamma$, where $\Gamma=2\pi|g|^{2}D(\varepsilon)$, where $D(\varepsilon)$ is density of states of atoms at energy $\varepsilon$ in the atomic bath. Throughout the paper, chemical potential of the source $\mu_{L}$ and the drain $\mu_{R}$ are assumed to be the same, as $\mu_{L}=\mu_{R}=\mu$, which are reflected by the Fermi-Dirac distribution functions $f_{\alpha}(\varepsilon_{s})=\frac{1}{e^{(\epsilon_{s}-\mu_{\alpha})/k_{B}T}+1}$. Here, $\mu$ indicates the equal quantity of these chemical potentials. $k_{B}$ is the Boltzmann constant and $T$ represents temperature of the two atomic baths.

Due to the spin-orbit coupling in atomtronics, momentum gain of atoms depends on the internal states of atoms. As a result, the ground state atoms and excited state atoms move in opposite directions. Using the relation of particle number change and input output current $\frac{d (\langle n_{1g}\rangle+\langle n_{1e}\rangle+\langle n_{2g}\rangle+\langle n_{2e} \rangle)}{d t}=I_{Lg}+I_{Le}-I_{Rg}-I_{Re}$~\cite{Hershfield,Davies}, we can reach the ground state atom current $
I_{g}=(I_{Lg}+I_{Rg})/2$ and the excited state atom current $I_{e}=(I_{Le}+I_{Re})/2$, where $I_{\alpha s}=\Gamma (f_{\alpha}(\varepsilon_{s})\langle a^{\dag}_{js}\rho a_{js}\rangle-(1-f_{\alpha}(\varepsilon_{s}))\langle\rho a^{\dag}_{js} a_{js}\rangle)$ with $j=1$ ($j=2$) when $\alpha=L$ ($\alpha=R$). In the formula $\langle n_{1g}\rangle+\langle n_{1e}\rangle+\langle n_{2g}\rangle+\langle n_{2e} \rangle$, the bra and ket represent average that calculated through the density matrix which is solved from the Eq.\eqref{equation}~\cite{average-atom}. The net atomic current can be obtained through $I=I_{g}+I_{e}$. As basic parameters, we take $\Omega=2\pi\times 600 Hz$, $\gamma=2\pi\times 500 Hz$, $\Gamma=2\pi\times 400 Hz$, $\Delta/\hbar=0.5\Omega$ and $k_{B}T/\hbar=0.1\Gamma$, which are based on the recent experiment \cite{Livi}.

\begin{figure}
\includegraphics[width=0.48\textwidth, clip]{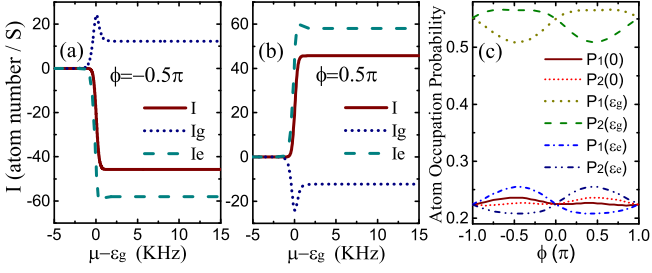}\\
\caption{(a)(b) Atomic net current $I$, ground state current $I_{g}$ and excited state current $I_{e}$ versus chemical
potential of the atom baths $\mu$. (c) Occupation probabilities of an atom in the first quantum dot
$P_{1}(0), P_{1}(\varepsilon_{g}), P_{1}(\varepsilon_{e})$ and in the second quantum dot $P_{2}(0), P_{2}(\varepsilon_{g}), P_{2}(\varepsilon_{e})$.}%
\label{ppp}
\end{figure}

\begin{figure}
\includegraphics[width=0.48\textwidth, clip]{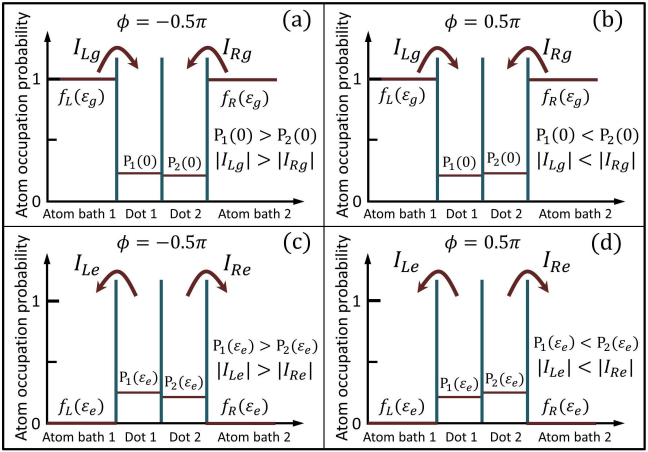}\\
\caption{(Color on line) At particular phases $\phi=-0.5\pi$ and $\phi=0.5\pi$, probability distributions of empty dots and excited atom occupations in optical potentials. Corresponding atomic current directions are shown using the arrows.}
\label{sysc}
\end{figure}

Energy levels of the model can be seen in Fig.~\ref{sys}(a). Only the ground states of the atomic quantum dots are considered and corresponding energies are set to be zero. Therefore, the energy in the quantum dots is represented by the ground energy $\varepsilon_{g}$ of atoms. If the chemical potential $\mu$ is lower than the ground energy of the double-dot, $\mu<\varepsilon_{g}$, the two quantum dots would always be empty and atomic current is zero as shown in Figs.~\ref{ppp}(a) and (b). When $\mu>\varepsilon_{e}$, then each dot is occupied by an atom instantaneously, at the same time, no atom can get into or out from the double dot system due to atom blocking effect. In this case, stable atom flow also can not occur.

Net atom current can be seen when the chemical potential $\mu$ is located between the ground level and excited level of the double-dot system, that is $\varepsilon_{g}<\mu<\varepsilon_{e}$ (see Fig.~\ref{sys}). It is the feature of cell which has effective voltage $\varepsilon_{e}-\varepsilon_{g}$ and current can be observed as soon as the potentials (for example $\mu$) of conductors are included in the energy range from $\varepsilon_{g}$ to $\varepsilon_{e}$. It is different from the quantum transport through quantum dots that driven by bias voltage, where at least one energy level of quantum dots must be located between the chemical potentials of source and drain. Since all the chemical potentials are the same, atom flow due to clock field transition is in fact a superfluid. Indeed, spin-orbit coupling can induce superfluid in Fermion atoms\cite{HHu,Jiang2011}. However, our system is an effective cell which has two poles (similar to electrodes) on the left and right side, respectively. Ground and excited state of atoms can be seen as two kinds of charges.

The atom occupation probabilities with , ground state, excited state and empty dot are defined as $P_{j}(0)=\langle \rho a_{js}a^{\dag}_{js}\rangle$, $P_{j}(\varepsilon_{g})=\langle\rho a^{\dag}_{jg} a_{jg}\rangle$ and $P_{j}(\varepsilon_{e})=\langle\rho a^{\dag}_{je} a_{je}\rangle$ with $j=1,2$. They satisfy $P_{j}(0)+P_{j}(\varepsilon_{g})+P_{j}(\varepsilon_{e})=1$. When chemical potential $\mu$ is much higher then the ground state level $\varepsilon_{g}$ as shown in Fig.~\ref{ppp} (a) \textbf{and} (b), the Fermion distribution functions are close to $1$ at the ground state level $\varepsilon_{g}$. We take a approximation that $f_{L}(\varepsilon_{g})=1$ and $f_{R}(\varepsilon_{g})=1$ and simplify the current $I_{g}$ to be
\begin{eqnarray}
I_{g}=\frac{\Gamma}{2}(P_{1}(0)-P_{2}(0)).
\label{simcurrentg}\end{eqnarray}
From Fig.~\ref{ppp}(c) and Fig~\ref{sysc} we know that, for the magnetic flux $-\pi<\phi<0$,the quantum dot occupation probabilities satisfy $P_{1}(0)>P_{2}(0)$. It leads to $|I_{Lg}|>|I_{Rg}|$ and furthermore $I_{g}>0$. For the magnetic flux $0<\phi<\pi$, the quantum dot occupation probabilities satisfy $P_{1}(0)<P_{2}(0)$. It gives the opposite result $|I_{Lg}|<|I_{Rg}|$ and $I_{g}<0$.

\begin{figure}
\includegraphics[width=0.48\textwidth, clip]{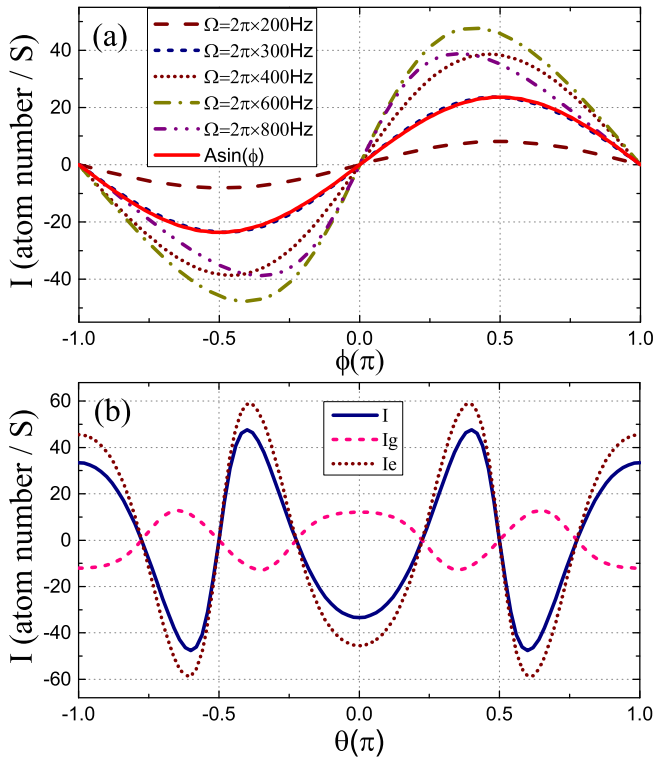}\\
\caption{(a) Atomic current as a function of the artificial magnetic flux $\phi$ at different Rabi frequencies.
$A sin\phi$ is a Sin function with amplitude $A$ (Here, $A=23.6$). (b) Atomic net current $I$, ground state current $I_{g}$ and excited state current $I_{e}$ as a function of the laser direction $|\theta|$.}
\label{ipitheta}
\end{figure}

The Fermion distribution functions of atom baths at the excited level $\varepsilon_{e}$ are nearly equal to $0$. Therefore, in the same way as above, we can take $f_{L}(\varepsilon_{e})=0$ and $f_{R}(\varepsilon_{e})=0$ to simplify the current $I_{e}$ in the form
\begin{eqnarray}
I_{e}=-\frac{\Gamma}{2}(P_{1}(\varepsilon_{e})-P_{2}(\varepsilon_{e})).
\label{simcurrente}\end{eqnarray}
Combining Eqs.\eqref{simcurrentg} and \eqref{simcurrente} with the numerical results in Figs.~\ref{ppp}(c), one can estimate that directions of ground state current and excited state current are always opposite. The input and output flows of atoms from the two baths are directly decided by the double-dot system. Eqs.\eqref{simcurrentg} and \eqref{simcurrente} indicate that the atom currents are proportional to the polarization of atom occupation probabilities. Different probabilities of the atom occupation in the two quantum dots can be represented by the phase difference of the atom wave function. Then the magnetic flux represents phase difference of atom wave function between the two quantum dots. Magnetic flux represents the phase difference of an atom wave function between the two quantum dots. Therefore, the artificial magnetic flux creates the atom currents. When the magnetic flux is integer times of $\pi$, the polarization of atom distribution probabilities would be disappear due to the periodic wave property of the atoms, and the current should be zero $I=0$ (see Fig.~\ref{ipitheta}(a)).

It is interesting that, at low Rabi frequency, the relation between atom current $I$ and phase difference $\phi$ almost satisfy the $Sin$ function as illustrated in Fig.~\ref{ipitheta}(a). In a good approximation, we can write it as $I=I_{0}sin \phi$ with a particular constant $I_{0}$ current. It is similar to the behavior of superconductor current in Josephson junction. The Josephson effect in cold atom is also predicted previously in Fermi superfluid~\cite{Jiang2016} and momentum space~\cite{Hou}. Because higher Rabi frequency makes the system become more sensitive to the artificial magnetic flux, the maximum point of the current in Fig.~\ref{ipitheta}(a) moves towards the center at high Rabi frequency.

\begin{figure}
\includegraphics[width=0.48\textwidth, clip]{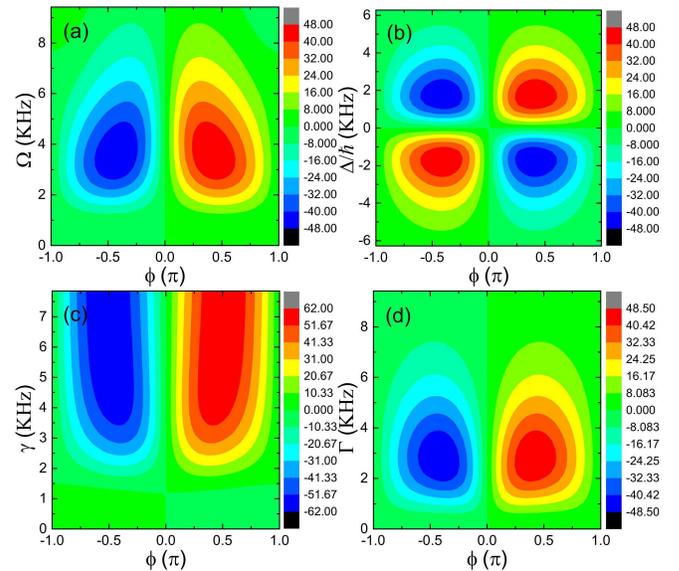}\\
\caption{Atomic net current $I$ versus (a) Rabi frequency $\Omega$, (b) laser \textbf{detuning} $\Delta$, (c) tunneling rate $\gamma$, and (d) coupling strength $\Gamma$ to the environment, along with the phase $\phi$ of magnetic flux. }%
\label{dogg}%
\end{figure}

For particular lasers with stationary wave length, the artificial magnetic flux $\phi$ only depends on the input angle $\theta$ of clock laser through the relation $\phi\propto \cos\theta$. Therefore, one can set an experiment as illustrated in Fig.~\ref{sys}, to control the atom current by changing the direction of incident clock laser. Amplitude and direction of current change along the variation of the angle $\theta$ with periodicity $2\pi$ as revealed in Fig.~\ref{ipitheta}(b). The current lines are mirror symmetry for the incident clock field moves clockwise and anticlockwise.

Fig.~\ref{dogg} (a) further certifies that the applied clock laser creates atomic current since $\Omega$ presents the atom-light coupling strength. Red-blue detuning determines the direction of atom flow due to the fact that energy loss and gain depends on the sign of $\Delta$ (See Fig.~\ref{dogg}(b)). The strong coupling between two quantum dots is propitious to coherent interaction of the system. Therefore, Fig.~\ref{dogg}(c) illustrates that increase of the tunneling rate $\gamma$ enhances atomic current. A proper large dot-bath coupling $\Gamma$ is needed for the occurrence of net current as illustrated in Fig.~\ref{dogg} (d), which emphasizes that an opened system is necessary for the photovoltaic transistor.

Atoms in ground state and excited state move in opposite directions, atom currents should be detectable at the two sides of the system through absorption and emission optical band in experiment~\cite{Goryca,Gall2010}. Optical clock transition which has longer coherent time comparing with the two-photon Raman transition~\cite{Dalibard} in which heat effect is unavoidable and the life time would be limited. Life-time of the optical clock transitions in alkaline-earth atoms or lanthanide atoms reach from $10$ $s$ to $10^{3}$ $s$~\cite{Nicholson12,Hinkley,Bloom,Nicholson15,Huang,Marti}, even at finite temperature.

A photovoltaic effect of single two-level atoms is illustrated using artificial gauge field assisted two atomic quantum dots. In the double quantum dots system, polarization of atom occupation probability is predicted which is basic feature of photovoltaic transistors. Current of the transistor is at single-atom level, which is important for few-atom manipulation. The atomic current can be controlled changing the direction of applied clock field or other system parameters. Previous photovoltaic cells are commonly for charged particles and we start a new platform of photovoltaic transistor that for neutral particles. Therefore, as a sub-micrometer sized basic transistor our system is scalable and stronger light induced current effect should be observed, which reveals its potential applications in neutral particle devices such as atom light sensor, atom flow amplifier and single atom transistor.

\begin{acknowledgments}
This work was supported by the National Key R and D Program of China under grants No. 2016YFA0301500, NSFC under grants Nos. 11434015,  61835013, the Strategic Priority Research Program of the Chinese Academy of Sciences under grants Nos. XDB01020300, XDB21030300. It was also supported by the NSF of Beijing under Grant No. 1173011, the Scientific Research Project of BMEC under Grant No. KM201711232019, and the Qin Xin Talents Cultivation Program of BISTU under Grant No. QXTCP C201711.
\end{acknowledgments}


\begin{thebibliography}{99}
\bibitem{average-atom} The average atom occupation numbers are defined as $\langle n_{1g}\rangle=Tr[\rho a^{\dag}_{1g}a_{1g}]$, $\langle n_{1e}\rangle=Tr[\rho a^{\dag}_{1e}a_{1e}]$, $\langle n_{2g}\rangle=Tr[\rho a^{\dag}_{2g}a_{2g}]$ and $\langle n_{2e}\rangle=Tr[\rho a^{\dag}_{2e}a_{2e}]$.
\end{thebibliography}
\end{document}